\documentclass{PoS}

\usepackage{amsmath}
\usepackage{mathrsfs}
\usepackage{cite}
\usepackage{tikz}

\newcommand{\dd}{\partial}
\renewcommand{\d}{\partial}
\newcommand{\half}{\frac{1}{2}}
\newcommand{\ffrac}[2]{\raisebox{.5pt}%
  {\footnotesize$\displaystyle\frac{#1}{#2}$}\kern1pt}

\newcommand{\ddl}[2]{\ffrac{\dd #1}{\dd #2}}

\newcommand{\vddl}[2]{{\ffrac{\delta #1}{\delta #2}}}

\title{Conserved currents in the Palatini formulation of general relativity}

\ShortTitle{Currents in the Palatini formulation of gravity}

\author{\speaker{Glenn Barnich}\\
        Physique Th\'eorique et Math\'ematique \\ Universit\'e libre de
   Bruxelles and International Solvay Institutes \\ Campus
   Plaine C.P. 231, B-1050 Bruxelles, Belgium\\
        E-mail: \email{gbarnich@ulb.ac.be}}

\author{Pujian Mao\\
        Center for Joint Quantum Studies and Department of Physics\\
     School of Science, Tianjin University\\
     135 Yaguan Road, Tianjin 300350, P. R. China\\
        E-mail: \email{pjmao@tju.edu.cn}}

\author{Romain Ruzziconi\\
        Physique Th\'eorique et Math\'ematique \\ Universit\'e libre de
   Bruxelles and International Solvay Institutes \\ Campus
   Plaine C.P. 231, B-1050 Bruxelles, Belgium\\
       E-mail: \email{rruzzico@ulb.ac.be}}

     \abstract{We derive the expressions for the local, on-shell
       closed co-dimension 2 forms in the Palatini formulation of
       general relativity and explicitly show their on-shell
       equivalence to those of the metric formulation. When compared
       to other first order formulations, two subtleties have to be
       addressed during the construction: off-shell non-metricity and
       the fact that the transformation of the connection under
       infinitesimal diffeomorphisms involves second order derivatives
       of the associated vector fields.}

     \FullConference{Corfu Summer Institute 2019 "School and Workshops
       on Elementary Particle Physics and Gravity" (CORFU2019)\\
       31 August - 25 September 2019\\
       Corfu, Greece}

\begin{document}

\section{Introduction}

The Palatini first order formulation is a convenient starting point
for the standard Hamiltonian approach to general relativity by
Arnowitt, Deser and Misner
\cite{Deser:1959zza,Arnowitt:1959ah,Arnowitt1960}. It is in this
framework that appropriate surface integrals at spatial infinity for
energy-momentum have originally been constructed
\cite{Arnowitt1960c,R.ArnowittS.DeserC.W.Misner1961} and that the
Hamiltonian formulation is presented in
\cite{Arnowitt:1962aa,Misner:1970aa}.

Conserved quantities in first order formulations of general relativity
have recently been investigated from a variety of perspectives, see
e.g.~\cite{Hehl:1994ue,Julia:1998ys,Julia:2000er,Julia:2002df,%
  Ashtekar:2008jw,JacobsonMohd2015,CorichiRubalcavaVukasinac2014,%
  Lehner:2016vdi,Korovin:2017xqu,DePaoli:2018erh,Oliveri:2019gvm}. In
the approach that we follow here
\cite{Barnich:2001jy,Barnich:2003xg,Barnich:2007bf}, one constructs
conserved co-dimension $2$ forms in the linearized theory from the
weakly vanishing Noether currents associated to gauge
symmetries. Indeed, one can show in the linearized theory that there
are conserved co-dimension $2$ forms for each reducibility parameter
of the background. The latter correspond to the Killing vectors of the
background metric in general relativity and one can show that there
are no other conserved co-dimension $2$ forms which are non-trivial.
The method has been applied recently to first order formulations of
general relativity where the variables are either a vielbein and a
Lorentz connection in coordinate basis \cite{Barnich:2016rwk}, or a
vielbein and the spin coefficients of the Newman-Penrose formalism
\cite{Barnich:2019vzx}. Two additional general results have been added
in that context: a general expression for conserved co-dimension $2$
forms applicable in a generic first order theory and a detailed
discussion of the breaking term, the flux terms that appear on the
right hand side of what would be a conservation law when one uses
general gauge parameters rather than reducibility parameters of the
background.

Two subtleties have to be faced when applying this construction to the
Palatini formulation of general relativity. The first is that the
theory is not first order in the sense that the transformation of the
connection under infinitesimal diffeomorphisms involves second order
derivatives of the vector field parametrizing these
diffeomorphisms. This leads to a weakly vanishing Noether current that
is not first order. It turns out however that all higher order terms
are contained in a total derivative and such terms are easily handled
by the contracting homotopy operator used to built the co-dimension
$2$ forms. As a consequence, the construction is as straightforward as
in other first order approaches to general relativity. The second
subtlety is that, as for other discussions of symmetries on the level
of an action principle, all computations are performed off-shell. For
the Palatini formalism, this means that one has to deal with
non-metricity.

The paper is organized as follows. We start with a very brief review
of how to construct conserved co-dimension 2 forms out of weakly
vanishing Noether currents. More details can be found in the original
literature cited above and an extensive recent summary has been
provided in \cite{Barnich:2019vzx}. We then discuss various identities
satisfied by the curvature tensor in the general context of a
non-holonomic frame including torsion and non-metricity because these
are relevant for the Noether identities that are crucial to the
construction. We then apply these general considerations to the
particular case of the Palatini formulation for which one uses a
coordinate basis and a connection without torsion. Finally we
construct the co-dimension $2$ forms and the associated breaking
terms.

\section{Construction of co-dimension 2 forms}

\subsection{General case}
\label{sec:expl-constr}

We consider a theory with a Lagrangian $n$-form
${\mathcal L}=L\, d^nx$ in $n$ dimensional spacetime. The fields of
the variational principle are denoted by $\phi^i$. Consider a
generating set (see e.g.~\cite{Henneaux:1992ig}, chapter 3) of non
trivial gauge transformations
$\delta_\epsilon \phi^i=R^i_\alpha(\epsilon^\alpha)$. One can prove
that there is an isomorphism between equivalence classes of on-shell
closed co-dimension $2$ forms and equivalence classes of reducibility
parameters $\bar f^\alpha[x,\phi]$ satisfying
$R^i_\alpha(\bar f^\alpha)\approx 0$. Equivalent co-dimension $2$
forms differ on-shell by an exact local form while equivalent sets of
reducibility parameters agree on-shell.

The relation between on-shell closed co-dimension $2$ forms and
reducibility parameters is constructive. For arbitrary gauge
parameters $f^\alpha$, a direct application of the Leibniz rule for
total derivatives leads to
\begin{equation}
  R^i_\alpha(f^\alpha)\vddl{\mathcal{L}}{\phi^i}=f^\alpha
  R^{+i}_\alpha\left(\vddl{\mathcal{L}}{\phi^i}\right)
  +d_H S_f,
\label{eq:1}
\end{equation}
for some weakly vanishing $n-1$ form
\begin{equation}
  S_f=S^{i\mu}_\alpha\left(\ddl{}{dx^\mu}\vddl{\mathcal{L}}{\phi^i},f^\alpha\right
  ). \label{eq:4} 
\end{equation}
The $n-2$ form can be constructed by using the contracting homotopy
$\rho_H$ for the horizontal differential of the variational bi-complex
\cite{Andersonbook,Olver:1993}
\begin{equation}
  \label{eq:5}
  \{d_H,\rho_H\}\omega^p=\omega^p\ {\rm for}\ p<n.
\end{equation}
Indeed, the Noether identities that are associated to the generating
set of non-trivial gauge transformations are
\begin{equation}
  \label{eq:6}
  R^{+i}_\alpha\left(\vddl{\mathcal{L}}{\phi^i}\right)=0.
\end{equation}
For reducibility parameters $\bar f^\alpha$, \eqref{eq:1} reduces to
$d_H S_{\bar f}\approx 0$. One then shows (see section 3.3 of
\cite{Barnich:2001jy} for details) that the weakly vanishing terms on
the right hand side can be absorbed by the horizontal differential of
a ``doubly'' weakly vanishing $n-1$ form $M_{\bar f}$ on the left hand
side, leading to $d_H (S_{\bar f}+M_{\bar f})=0$. When applying the
contracting homotopy to $J_{\bar f}=S_{\bar f}+M_{\bar f}$,
\begin{equation}
  \label{eq:7}
  k_{\bar f}=\rho_H J_{\bar f},
\end{equation}
it then follows from \eqref{eq:5} that 
\begin{equation}
  \label{eq:8}
  d_H k_{\bar f}= J_{\bar f}\approx 0.
\end{equation}
In case where one can show that a set of reducibility parameters is
equivalent to a set for which $R^i_\alpha(\bar f^\alpha) =0$, the
reasoning simplifies since in this case $d_H S_{\bar f}= 0$, and the
application of \eqref{eq:5} now directly yields
$d_H k_{\bar f}= S_{\bar f}\approx 0$ with
$k_{\bar f}=\rho_H S_{\bar f}$. Similarily, in linear gauge theories,
the application of the homotopy formula to $M_{\bar f}$ gives rise to
a weakly vanishing and thus trivial $n-2$ form, which can be
omitted. In this case, we still have $k_{\bar f}=\rho_H S_{\bar f}$
but now $d_H k_{\bar f}\approx S_{\bar f}\approx 0$.

\subsection{Linearized theories and asymptotics}
\label{sec:linearized-theories}

For the purposes of exposition, we focus on the Einstein-Hilbert
action in metric formulation, where a generating set of gauge
transformations corresponds to the Lie derivative of the metric,
$\delta_\xi g_{\mu\nu}={\mathcal L}_\xi g_{\mu\nu}$. In spacetime
dimensions $n\geq 3$, one can then show that all equivalence classes
of reducibility parameters admit representatives $\xi^\rho[x]$ that do
not depend on $g_{\mu\nu}$ and its derivatives. The condition that
such vectors are reducibility parameters then reduces to the Killing
equation for a generic metric. Since a generic metric does not have
Killing vectors, no non-trivial conserved $n-2$ forms can be
constructed in general relativity. However, one can linearize the
theory around a background solution $\bar g_{\mu\nu}$. A generating
set of gauge transformations of the linearized theory corresponds to
the Lie derivative of the background metric,
$\delta_\xi h_{\mu\nu}={\mathcal L}_\xi\bar g_{\mu\nu}$. It then
follows that there are as many conserved $n-2$ forms as there are
Killing vectors of the background solution. The explicit expressions
of the $n-2$ forms are obtained by applying the construction described
in previous subsection in the framework of the linearized theory. This
has been done explicitly for Einstein gravity in
\cite{Barnich:2001jy}.

More generally, one can show \cite{Barnich:2003xg} that the $n-2$
forms of the linearized theory can be obtained from the weakly
vanishing co-dimension 1 form $S_f$ of the full theory through
\begin{equation}
  \label{eq:9}
  k_f[\delta\phi,\phi]=k^{\mu\nu}_f(d^{n-2}x)_{\mu\nu}=\frac{|\lambda|+1}{|\lambda|+2}
\partial_{(\lambda)}\left[\delta\phi^i\vddl{}{\phi^i_{((\lambda)\nu)}}\ddl{}{dx^\nu}
S_f\right],
\end{equation}
by replacing $f$ by reducibility parameters of the linearized theory,
$\phi^i$ by the background solution $\bar\phi^i$ and $\delta\phi^i$ by
any solution $\bar\varphi^i$ of the theory linearized around
$\bar\phi^i$. We refer to \cite{Andersonbook} and \cite{Olver:1993}
for the explicit expressions for the higher order Euler-Lagrange
derivatives. Our conventions and notations for multi-indices are
summarized in the appendix of \cite{Barnich:2001jy}.

For theories such as general relativity in metric formulation where
$S_f$ is at most of second order in derivatives, the formula involves
the higher order Euler-Lagrange operators only up to order $2$ and
reduces to 
\begin{equation}
  \label{eq:64}
  k_f[\delta\phi,\phi]=\frac{1}{2}\delta\phi^i\vddl{}{\phi^i_\nu}\ddl{}{dx^\nu}
  S_f+\frac{2}{3}\d_\sigma\left[\delta\phi^i\vddl{}{\phi^i_{\nu\sigma}}\ddl{}{dx^\nu}
    S_f\right].
\end{equation}
For later use, note that for a local function $M$ involving the fields
and their derivatives up to some finite order, a key property of the
higher order Euler-Lagrange derivatives is that they ``absorb'' total
derivatives,
\begin{equation}
  \frac{\delta \partial_\lambda M }{\delta \phi^i}=0,\qquad
  \frac{\delta \partial_\lambda M }{\delta \phi^i_\nu}
  = \delta^\nu_\lambda \frac{\delta  M }{\delta \phi^i}, \qquad
  \frac{\delta \partial_\lambda M}{\delta \phi^i_{\mu \nu}}
  = \delta^{(\mu}_\lambda \frac{\delta M}{\delta \phi^i_{\nu)}},
\label{useful identities M}
\end{equation}
where the round (square) brackets denote (anti) symmetrization of
enclosed indices divided by the factorial of the number of indices
involved. Furthermore, if $M^1$ depends at most on first order
derivatives, the Euler-Lagrange derivatives of order
one reduce to partial derivatives, 
\begin{equation}
\frac{\delta M^1}{\delta
  \phi^i_\nu}=\frac{\d M^1}{\d \phi^i_\nu}.\label{eq:21}
\end{equation}

As shown in detail in \cite{Barnich:2019vzx}, when using general gauge
parameters $f^\alpha$ instead of reducibility parameters,
non-conservation is controlled by the co-dimension $1$ form
$b[\delta\phi,R_f,\phi]$ defined by
\begin{equation}
  \label{eq:10}
  b=-\d_{(\lambda)}\left[R^i_{\alpha}(f^\alpha)
  \delta\phi^j\frac{\delta}{\delta
    \phi^j_{(\lambda)\nu}}\frac{\partial}{\partial dx^\nu}\left(\frac{\delta
      \mathcal L}{\delta \phi^i}\right)\right]
\end{equation}
which satisfies $b[\delta\phi,R_f,\phi]=-b[R_f,\delta\phi,\phi]$ by
construction.  Indeed, when $\phi^i$ is a solution to the equations of
motion, $\delta\phi^i$ a solution to the linearized equations of
motion, the co-dimension 2 form $k_f$ constructed as in \eqref{eq:9}
is is no longer $d_H$-closed but satisfies instead
\begin{equation}
  \label{eq:3}
  d_H k_f=b. 
\end{equation}
As in asymptotically flat general relativity at null infinity
\cite{Wald:1999wa,Barnich:2011mi,Barnich:2013axa}, these on-shell
non-closed co-dimension 2 forms $k_f$ are in general not integrable
either.

\section{Vielbeins and connection}
\label{sec:first}

Now, we recall some notions of vielbeins and connection by including
torsion and non-metricity into the standard discussion. In particular,
this completes the results of \cite{Barnich:2016rwk,Barnich:2019vzx}
by considering non-metricity.

\subsection{General case}
\label{sec:general-case}

Consider an $n$-dimensional spacetime with a moving frame (or
vielbein)
\begin{equation}
e_a={e_a}^\mu\ddl{}{x^\mu},\quad e^a={e^a}_\mu dx^\mu, \label{eq:2}
\end{equation}
where ${e_a}^\mu{e^a}_\nu=\delta^\mu_\nu$,
${e_a}^\mu{e^b}_\mu=\delta_a^b$, and $\d_a f=e_a(f)$. The structure functions are defined by
\begin{equation}
[e_a,e_b]={D^c}_{ab}e_c \iff de^a=-\half {D^a}_{bc}e^be^c.\label{eq:13}
\end{equation}
For further use, note that if ${\mathbf e}={\rm det}\,{e^a}_\mu$, then
\begin{equation}
  \label{eq:82}
  \d_\mu(\mathbf{e}\,{e^\mu}_a)=\mathbf{e}\, {D^b}_{ba},
\end{equation}
and, if we define,
\begin{equation}
  \label{eq:49}
   {d^a}_{bc}={e^a}_\lambda \d_b {e_c}^\lambda,
\end{equation}
then
\begin{equation}
{d^\sigma}_{\rho\mu}=-{e_d}^\sigma\d_\rho {e^d}_\mu,\quad
{D^a}_{bc}=2{d^a}_{[bc]}, \label{eq:100}
\end{equation}
where it is understood that tangent space indices $a,b,\dots$ and
world-indices $\mu,\nu,\dots$ are transformed into each other by using
the vielbeins and their inverse.

In addition, we assume that there is an affine connection
\begin{equation}
D_a e_b={\Gamma^c}_{ba}e_c\iff D_b v^a=\d_b v^a+{\Gamma^a}_{cb} v^c. \label{eq:12}
\end{equation}
The components of the torsion tensor are given by
\begin{equation}
{T^a}_{\mu\nu}=\d_\mu {e^a}_\nu-\d_\nu
{e^a}_\mu+{\Gamma^a}_{b\mu}
{e^b}_\nu-{\Gamma^a}_{b\nu}{e^b}_\mu,\label{eq:70}
\end{equation}
\begin{equation}
{T^c}_{ab}=2{\Gamma^c}_{[ba]}+{D^c}_{ba}=2({\Gamma^c}_{[ba]}+{d^c}_{[ba]}),\label{eq:14}
\end{equation} while the components of the curvature tensor can be written as
\begin{equation}
{R^f}_{c\mu\nu}=\d_\mu
  {\Gamma^f}_{c\nu}-\d_\nu {\Gamma^f}_{c\mu}
  +{\Gamma^f}_{d\mu}{\Gamma^d}_{c\nu}-{\Gamma^f}_{d\nu}{\Gamma^d}_{c\mu},\label{eq:71}
\end{equation}
\begin{equation}
{R^f}_{cab}=\d_a
  {\Gamma^f}_{cb}-\d_b {\Gamma^f}_{ca}
  +{\Gamma^f}_{da}{\Gamma^d}_{cb}-{\Gamma^f}_{db}{\Gamma^d}_{ca}-{D^d}_{ab}{\Gamma^f}_{cd}.
\label{eq:15}
\end{equation}
Furthermore,
\begin{equation}
  \label{eq:20}
  [D_a,D_b]v_c=-{R^d}_{cab}v_d-{T^d}_{ab}D_dv_c.
\end{equation}
The Bianchi
identities are given explicitly by
\begin{equation}
  \label{eq:24}
  {R^a}_{[bcd]}=D_{[b}{T^a}_{cd]}+{T^a}_{f[b}{T^f}_{cd]},\quad
D_{[f}{R^a}_{|b|cd]}=-{R^a}_{bg[f}{T^g}_{cd]},
\end{equation}
where a bar encloses indices that are not involved in
the (anti) symmetrization. The Ricci tensor is defined by
${R}_{ab}={R^{c}}_{acb}$, while
$S_{ab}={R^c}_{cab}$. Contracting the Bianchi identities gives
\begin{equation}
  \label{eq:27}
  {R}_{ab}-{R}_{ba}=S_{ab}-D_c {T^c}_{ab}
-2D_{[a} {T^c}_{b]c}-{T^c}_{dc}{T^d}_{ab},
\end{equation}
\begin{equation}
2D_{[f}{R}_{|b|d]}+D_c{R^c}_{bdf}={R}_{bg}{T^g}_{df}
-2{R^c}_{b[f|g|}{T^g}_{d]c}, \label{eq:28a}
\end{equation}
\begin{equation}
  \label{eq:28b}
  D_{[f}S_{cd]}=-S_{g[f}{T^g}_{cd]}.
\end{equation}

Assume now that there is a pseudo-Riemannian metric,
\begin{equation}
g_{\mu\nu}={e^a}_\mu g_{ab} {e^b}_\nu\label{eq:11},
\end{equation}
i.e., a symmetric, non-degenerate $2$-tensor.
As usual, tangent space indices $a,b,\dots$ and world indices
$\mu,\nu,\dots$ are lowered and raised with $g_{ab}$, $g_{\mu\nu}$,
and their inverses.
The non-metricity tensor is defined as
$\Xi^{ab}=dg^{ab}+2\Gamma^{(ab)}$. The associated Bianchi identities
are given by
$d\Xi^{ab}+{\Gamma^{a}}_c\Xi^{cb}+{\Gamma^b}_c\Xi^{ac}=2R^{(ab)}$.
More explicitly,
\begin{equation}
  \label{eq:95}
  {\Xi^{ab}}_c=D_c g^{ab}, \quad 2D_{[c}
  {\Xi^{ab}}_{d]}=-{\Xi^{ab}}_f{T^f}_{cd}+2{R^{(ab)}}_{cd}.
\end{equation}
Note
also that, from $g^{ab}g_{bc}=\delta^a_c$, it follows that
\begin{equation}
  \label{eq:40}
  D_c g_{ab}=-\Xi_{abc}.
\end{equation}
Contracting the last of \eqref{eq:95} with $g_{ab}$
gives
\begin{equation}
  \label{eq:86}
  S_{cd}=g_{ab}D_{[c} {\Xi^{ab}}_{d]}+\half
  {\Xi^a}_{af}{T^f}_{cd},
\end{equation}
while \eqref{eq:28a} contracted with $g^{bf}$ gives
\begin{multline}
  \label{eq:34}
  D^b {R}_{ba}-\half D_a R=\half
  {R^{bc}}_{da}{T^{d}}_{bc}+{{R}^b}_c{T^c}_{ab}\\- \half({\Xi^{bc}}_c{
    R}_{ba}+{\Xi^{cd}}_b{R^b}_{cda}+{\Xi^{bc}}_a{
    R}_{bc})\\+D_c(D_{[b}{\Xi^{bc}}_{a]}+\half{\Xi^{bc}}_d{T^d}_{ba})+
  (D_{[b}{\Xi^{bc}}_{d]}+\half{\Xi^{bc}}_d{T^d}_{bd}){T^d}_{ac}.
\end{multline}

The curvature scalar is defined by
${R}=g^{ab}{R}_{ab}$, the Einstein tensor by
\begin{equation}
  \label{eq:33}
  G_{ab}={R}_{(ab)}-\half g_{ab} {R}.
\end{equation}
When combining with \eqref{eq:27}, the contracted Bianchi identity
\eqref{eq:34} written in terms of the Einstein tensor is
\begin{multline}
  \label{eq:34a}
D^b {G}_{ba}=\half
  {R^{bc}}_{da}{T^{d}}_{bc}+{{R}^b}_c{T^c}_{ab} - \frac{1}{2} {\Xi_{ab}}^b R   \\
 +\half D^b(S_{ab}-D_c {T^c}_{ab}
-2D_{[a} {T^c}_{b]c}-{T^c}_{dc}{T^d}_{ab})\\ - \half({\Xi^{bc}}_c{
    R}_{ba}+{\Xi^{cd}}_b{R^b}_{cda}+{\Xi^{bc}}_a{
    R}_{bc})\\+D_c(D_{[b}{\Xi^{bc}}_{a]}+\half{\Xi^{bc}}_d{T^d}_{ba})+
  (D_{[b}{\Xi^{bc}}_{d]}+\half{\Xi^{bc}}_d{T^d}_{bd}){T^d}_{ac} .
\end{multline}

By the usual manipulations, one may show in full generality that the existence of the metric
implies that the most general connection can be written as
\begin{equation}
  \label{eq:17}
  \Gamma_{abc}=
  \{{}_{abc}\}+M_{abc}+K_{abc}+r_{abc} ,
\end{equation}
where 
\begin{equation}
\{{}_{abc}\}=\half(g_{ab,c}+g_{ac,b}-g_{bc,a})=\{{}_{acb}\}, \label{eq:97a}
\end{equation}
\begin{equation}
M_{abc}=\half(\Xi_{abc}+\Xi_{acb}-\Xi_{bca})=M_{acb}, \label{eq:97b}
\end{equation}
\begin{equation}
  \label{eq:99c}
K_{abc}=\half(T_{bac}+T_{cab}-T_{abc})=-K_{bac},
\end{equation}
\begin{equation}
  r_{abc}=\half(D_{bac}+D_{cab}-D_{abc})=-r_{bac}.\label{eq:96}
\end{equation}
Furthermore, one can directly show that
\begin{equation}
  \label{eq:105}
  {\Gamma^a}_{b\mu}={e^a}_\nu(\d_\mu
  {e_b}^\nu+{\Gamma^\nu}_{\rho\mu}{e^\rho}_b)\iff
{\Gamma}_{abc}=e_{a\nu}\d_c{e_b}^\nu+{e_a}^\mu{e_b}^\nu{e_c}^\rho\Gamma_{\mu\nu\rho}.
\end{equation}
Finally, we will need the following variation
\begin{equation}
  \label{eq:47}
  \delta
  {R^a}_{b\mu\nu}=D_\mu\delta{\Gamma^a}_{b\nu}-D_\nu\delta{\Gamma^a}_{b\mu}.
\end{equation}

\subsection{Coordinate basis, torsionless connection}
\label{sec:coordinate-basis}

We now consider the particular case of a coordinate basis,
${e_a}^\mu={\delta_a}^\mu$ so that ${D^\lambda}_{\mu\nu}=0$ and
${T^\lambda}_{\mu\nu}={\Gamma^\lambda}_{\nu\mu}-{\Gamma^\lambda}_{\mu\nu}$. We
also impose vanishing of torsion, which requires the connection to be
symmetric, ${\Gamma^\lambda}_{\mu\nu}= {\Gamma^\lambda}_{\nu\mu}$. In
this case, equation \eqref{eq:27} implies
$S_{\mu\nu}=R_{\mu\nu}-R_{\nu\mu}$ and the contracted Bianchi
identities \eqref{eq:34a} become
\begin{equation}
  \label{eq:34c}
D^\nu {G}_{\nu\mu}=
 D^\nu R_{[\mu\nu]} +D_\lambda
 {R^{(\lambda\nu)}}_{\nu\mu}
 - \half(D_\nu g^{\nu\lambda}{
    R}_{\lambda\mu}+D_\nu g^{\lambda\rho} {R^\nu}_{\lambda\rho\mu}+
  D_\mu g^{\nu\lambda} {
    R}_{\nu\lambda}+ D^\nu g_{\mu \nu} R),
\end{equation}
while the variation \eqref{eq:47} simplifies to
\begin{equation}
  \label{eq:51}
   \delta
  {R^\alpha}_{\beta\mu\nu}=D_\mu\delta{\Gamma^\alpha}_{\beta\nu}
-D_\nu\delta{\Gamma^\alpha}_{\beta\mu}.
\end{equation}
We also have
\begin{equation}
\d_\mu(\sqrt{|g|} v^\mu)=\sqrt{|g|}
(D_\mu-{\Gamma^\nu}_{\mu\nu} +\half g^{\nu\lambda}\d_\mu
g_{\nu\lambda}) v^\mu
=D_{\mu}(\sqrt{|g|}v^\mu),
\label{eq:41}
\end{equation}
where the last equality follows by introducing the convenient
definition for the covariant derivative of a scalar density,
\begin{equation}
  \label{eq:53}
   D_\mu\sqrt{|g|}=\sqrt{|g|}(\half g^{\nu\lambda}\d_\mu
 g_{\nu\lambda}-{\Gamma^\nu}_{\mu\nu}).
 \end{equation}


If in addition, as will be imposed below on-shell, one requires
metricity, $\Xi^{ab}=dg^{ab}+2\Gamma^{(ab)} =0$, one recovers the
standard Christoffel connection
\begin{equation}
  \label{eq:18}
  \Gamma_{\lambda\mu\nu}=\half(\d_\nu g_{\lambda\mu}+\d_\mu
  g_{\lambda\nu}-\d_\lambda g_{\mu\nu}),
\end{equation}
The contracted Bianchi identities \eqref{eq:34c} reduce to
\begin{equation}
  \label{eq:42}
  D^\nu {G_{\nu\mu}}=0,
\end{equation}
and \eqref{eq:41} to
\begin{equation}
  \label{eq:43}
  \d_\mu(\sqrt{|g|} v^\mu)=\sqrt{|g|} D_\mu v^\mu.
\end{equation}

\section{Palatini formulation}
\label{sec:palatini-formulation}

\subsection{Variational principle}
\label{sec:vari-princ-Pal}

In the formulation discussed for example in \cite{Misner:1970aa}, one
uses a coordinate basis ${e_a}^\mu=\delta^\mu_a$,
${D^\mu}_{\nu\rho}=0$ with a metric $g_{\mu\nu}$ and a torsionfree
connection ${\Gamma^\lambda}_{\mu\nu}={\Gamma^\lambda}_{\nu\mu}$ as
variables\footnote{Adapting the arguments below to the case where the
  variables are chosen as the contravariant metric tensor density and
  the connection as done in \cite{Deser:1959zza,Arnowitt:1959ah} is
  straightforward.} to write the Palatini action as
\begin{equation}
  \label{eq:3P}
  S^P[g_{\mu\nu},{\Gamma^\lambda}_{\mu\nu}]=\kappa \int d^nx\, L^P=
  \kappa \int d^nx \sqrt {|g|} 
  ({ R}-2\Lambda),
\end{equation}
where $\kappa^{-1} = 16\pi G$ and we assume $n\geq 3$. Using \eqref{eq:51}, the variation of the
action is given by
\begin{equation}
  \label{eq:52}
  \delta S^P= \kappa \int d^nx \sqrt {|g|}\big[-(G^{\mu\nu}+\Lambda
  g^{\mu\nu})\delta
  g_{\mu\nu}+g^{\alpha\beta}(D_\mu\delta{\Gamma^\mu}_{\alpha\beta}
-D_\beta\delta{\Gamma^\mu}_{\alpha\mu})\big].
\end{equation}
Using in addition \eqref{eq:41} and neglecting boundary terms yields
\begin{multline}
  \label{eq:55}
  \delta S^P=\kappa \int d^nx \big[-\sqrt {|g|}(G^{\mu\nu}+\Lambda
  g^{\mu\nu})\delta
  g_{\mu\nu}\\+(-D_\mu[\sqrt {|g|}g^{\alpha\beta}]+D_\lambda[\sqrt
  {|g|}g^{\alpha\lambda}\delta_\mu^\beta])
  \delta{\Gamma^\mu}_{\alpha\beta}
  \big],
\end{multline}
so that the Euler-Lagrange derivatives of $L^P$ with respect to the
fields $g_{\mu\nu}$ and ${\Gamma^\lambda}_{\mu\nu}$ take the form
\begin{align}
  \label{eq:56}
  \vddl{L^P}{g_{\mu\nu}} &=-\sqrt {|g|}(G^{\mu\nu}+\Lambda
  g^{\mu\nu}), \\
  \vddl{L^P}{{\Gamma^\mu}_{\alpha\beta}} &= -D_\mu[\sqrt
  {|g|}g^{\alpha\beta}]+\half D_\lambda[\sqrt
  {|g|}g^{\alpha\lambda}\delta_\mu^\beta]+\half D_\lambda[\sqrt
  {|g|}g^{\beta\lambda}\delta_\mu^\alpha].\label{eq:57}
\end{align}
Contracting the equations of motion corresponding to \eqref{eq:57}
with $\delta^\mu_\beta$ gives
$D_\beta[\sqrt{|g|}g^{\alpha\beta}]=0$. When
re-injecting this result into the equation of motion, this implies $D_\mu[\sqrt{|g|}g^{\alpha\beta}]=0$. From
${\rm det}(\sqrt{|g|}g^{\alpha\beta})=|g|^{\frac{n-2}{2}}$, one then
deduces that
\begin{equation*}
  \delta |g|^\half=\delta ({\rm
    det}(\sqrt{|g|}g^{\alpha\beta}))^{\frac{1}{n-2}}= \frac{1}{n-2}
  {\rm det}(\sqrt{|g|}g^{\alpha\beta})^{\frac{1}{n-2}-1}\delta {\rm
    det}(\sqrt{|g|}g^{\alpha\beta})\\
=\frac{1}{n-2}g_{\alpha\beta}\delta(\sqrt{|g|}g^{\alpha\beta}).
\end{equation*}
When the variation corresponds to the covariant derivative $D_\mu$, we
deduce that these equations of motion imply that $D_\mu \sqrt{|g|}=0$,
and then metricity, $D_\mu g^{\alpha\beta}=0$. Since this implies
\eqref{eq:18}, it follows that ${\Gamma^\mu}_{\alpha\beta}$ are
auxiliary fields, i.e., fields that can be eliminated algebraically by
their own equations of motion. 

\subsection{Gauge symmetries and Noether identities}

If $\xi^\mu(x)$ denotes the vector field parametrizing an infinitesimal
diffeomorphism, the variation of the variables of the variational
principle at a given point is 
\begin{align}
  \delta_\xi g_{\mu\nu} &= \mathcal{L}_\xi g_{\mu\nu}
  = \xi^\rho \partial_\rho g_{\mu\nu} + g_{\rho \nu} \partial_\mu
                          \xi^\rho
                          + g_{\mu \rho} \partial_\nu
                          \xi^\rho, \label{variation g} \\ 
  \delta_\xi
  {\Gamma^\mu}_{\nu\rho}&=\d_\rho\d_\nu\xi^\mu+\xi^\sigma\partial_\sigma
                          {\Gamma^\mu}_{\nu\rho}-\d_\sigma\xi^\mu
                          {\Gamma^\sigma}_{\nu\rho}+\d_\nu\xi^\sigma
                          {\Gamma^\mu}_{\sigma\rho}+\d_\rho\xi^\sigma
                          {\Gamma^\mu}_{\nu\sigma}. \label{variation
                          Gamma} 
\end{align}
These transformations are infinitesimal gauge symmetries of the
Palatani formulation in the sense that
$\delta_\xi L^P=\d_\mu(\xi^\mu L^P)$ for all $\xi^\mu(x)$. As usual,
this follows as a consequence of the fact that $L^P$ transforms like a
scalar density under finite diffeomorphisms.

At this stage, we note that the transformation law of the connection
in \eqref{variation Gamma} involves derivatives of the gauge parameter
$\xi^\mu$ up to second order. Therefore, even though the
Euler-Lagrange equations \eqref{eq:56} and \eqref{eq:57} are of first
order, the theory is not in the class of first order theories
described for instance \cite{Barnich:2019vzx}.

The Noether identities and the weakly vanishing Noether current
associated to these gauge symmetries are then identified by using
Leibniz rule to write the analog of \eqref{eq:1} in the present
case,
\begin{equation}
  \label{eq:62}
 \vddl{\kappa L^P}{g_{\mu\nu}}\delta_\xi
  g_{\mu\nu}+\vddl{\kappa L^P}{{\Gamma^\mu}_{\alpha\beta}}\delta_\xi
  {\Gamma^\mu}_{\alpha\beta} =\xi^\rho N_\rho+\d_\mu
   S^\mu_\xi,
\end{equation}
which leads to the Noether identities
\begin{multline}
  \label{eq:54}
  \kappa^{-1} N_\rho=\vddl{L^P}{g_{\mu\nu}}\d_\rho g_{\mu\nu}-2\d_\sigma
  \left(\vddl{L^P}{g_{\sigma\nu}}g_{\rho\nu}\right)\\+
  \d_\alpha\d_\beta \left(\vddl{L^P}{{\Gamma^\rho}_{\alpha\beta}}
  \right)
  +\vddl{L^P}{{\Gamma^\mu}_{\alpha\beta}}\d_\rho{\Gamma^\mu}_{\alpha\beta}
  +\d_\sigma
  \left(\vddl{L^P}{{\Gamma^\rho}_{\alpha\beta}}{\Gamma^\sigma}_{\alpha\beta}
  \right) -2\d_\sigma
  \left(\vddl{L^P}{{\Gamma^\mu}_{\sigma\beta}}{\Gamma^\mu}_{\rho\beta}
  \right)=0.
\end{multline}
These identities correspond to the contracted Bianchi identities
\eqref{eq:34c}. Indeed, \eqref{eq:54} can be rewritten as
\begin{equation}
\label{Noeth Pal}
\vddl{L^P}{g_{\mu\nu}} D_\rho g_{\mu\nu} - 2 D_\mu \left( g_{\rho \nu}
  \vddl{L^P}{g_{\mu\nu}}
\right) + \vddl{L^P}{{\Gamma^\tau}_{\sigma\nu}} {R^\tau}_{\sigma \rho
  \nu}
+ D_\sigma D_\nu \left( \vddl{L^P}{{\Gamma^\rho}_{\sigma\nu}} \right) = 0.
\end{equation}
By inserting \eqref{eq:56} and \eqref{eq:57} into \eqref{Noeth Pal},
one recovers \eqref{eq:34c}. For this computation, the
identities 
\begin{equation}
[D_\mu , D_\nu ] \sqrt{|g|} = -  \sqrt{|g|} ({R}_{\mu \nu} - {R}_{\nu \mu}),
\end{equation}
\begin{equation}
[D_\mu , D_\nu ] D_\lambda \sqrt{|g|} = -  D_\tau \sqrt{|g|}
{R^\tau}_{\lambda \mu \nu} - D_\lambda \sqrt{|g|}({R}_{\mu \nu}
- {R}_{\nu \mu}), 
\end{equation}
are useful.

\subsection{Construction of the co-dimension 2 form}
\label{sec:constr-co-dimens-Pal}

We also get from \eqref{eq:62} the weakly vanishing Noether current
associated with the gauge symmetries,
\begin{equation}
\label{3formePalatini}
\kappa^{-1} S^\mu_\xi=2\frac{\delta L^P}{\delta g_{\mu\tau}}\xi_\tau +
2 \frac{\delta L^P}{\delta {\Gamma^\tau}_{\mu\rho}}D_\rho
\xi^\tau-{\Gamma^\mu}_{\rho\sigma}\frac{\delta L^P}{\delta
    {\Gamma^\tau}_{\sigma\rho}}\xi^\tau  - \partial_\rho
\left(\frac{\delta L^P}{\delta {\Gamma^\tau}_{\mu\rho}}\xi^\tau
\right). 
\end{equation}
In order to compute the co-dimension 2 forms, one now needs to insert
this expression into the general formula \eqref{eq:9}. Note that
\eqref{3formePalatini} involves second order derivatives of the fields
in the last term as a consequence of the second order derivatives on
the gauge parameters. Since these second derivatives occur under a
total derivative however, the properties \eqref{useful identities M}
allow one to reduce the actual computation to one involving only the
Euler-Lagrange derivatives of order one acting on expressions which
are at most of first order in derivatives, 
\begin{multline}
\label{2formePalatini}
\kappa^{-1} k^{[\mu\nu]}_\xi=\half\delta\phi^i\frac{\delta}{\delta
  \phi^i_\nu}
\left[2\frac{\delta L^P}{\delta g_{\mu\tau}}\xi_\tau
  + 2 \frac{\delta L^P}{\delta {\Gamma^\tau}_{\mu\rho}}D_\rho
  \xi^\tau-{\Gamma^\mu}_{\rho\sigma}\frac{\delta L^P}{\delta
      {\Gamma^\tau}_{\sigma\rho}}\xi^\tau  \right]\\
-\frac{1}{3}\partial_\rho \left[\delta\phi^i\frac{\delta}{\delta
    \phi^i_{\nu}}
  \left(\frac{\delta L^P}{\delta {\Gamma^\tau}_{\mu\rho}}\xi^\tau
  \right)\right]
-(\mu\leftrightarrow\nu).
\end{multline}
To proceed in this computation, let us introduce the notations
$h_{\mu\nu}=\delta g_{\mu\nu}$,
$\delta {\Gamma^\rho}_{\mu \nu} = {C^\rho}_{\mu \nu}$, indices being
lowered and raised with $g_{\mu\nu}$ and its inverse, and
$h=h^\mu_\mu$. Using
\begin{equation*}
  \begin{split}
& \delta ( \sqrt{|g|} g^{\mu \lambda} ) = - \sqrt{|g|} h^{\mu \lambda}
+ \half \sqrt{|g|} g^{\mu \lambda} h,\\
& \partial_\lambda ( \sqrt{|g|} h^{\mu \lambda} \xi^\nu) = D_\lambda (
\sqrt{|g|} h^{\mu \lambda} \xi^\nu) - \sqrt{|g|} {\Gamma^\mu}_{\lambda
  \tau} h^{\tau \lambda} \xi^\nu - \sqrt{|g|} {\Gamma^\nu}_{\lambda
  \tau} h^{\mu \lambda} \xi^{\tau},
\end{split}
\end{equation*}
one finally obtains the explicit
expression for the co-dimension 2 form, 
\begin{multline}
\label{Final expression Pal}
  \kappa^{-1} k^{[\mu\nu]}_\xi = \sqrt{|g|} \Big[\xi^\sigma
  C^{\mu\nu}_{\;\;\;\sigma}-\xi^\mu
  C_{\;\;\sigma}^{\sigma\;\;\;\nu}+\half\xi^\mu
  C^{\nu\sigma}_{\;\;\;\sigma}- h^{\nu\sigma}D_\sigma\xi^\mu +\half h
  D^\nu\xi^\mu \\ -\half \xi^\nu D_\sigma
  h^{\mu\sigma}+\frac{1}{4}\xi^\nu D^\mu h\Big] -\half
  h^{\mu\lambda}\xi^\nu D_\lambda( \sqrt{|g|})+\frac{1}{4}h\xi^\nu
  D_\lambda(\sqrt{|g|} g^{\mu\lambda})-(\mu\leftrightarrow\nu).
\end{multline}
The breaking term is easy to work out since it merely involves the
Euler-Lagrange derivatives of order one acting on expressions that are
at most of first order in derivatives. It is given by
\begin{equation}
  \kappa^{-1} b^\mu= \delta_\xi {\Gamma^\mu}_{\rho\nu} \delta (\sqrt {|g|} g^{\nu\rho} ) -
  \delta_\xi {\Gamma^\nu}_{\rho\nu} \delta (\sqrt {|g|} g^{\mu\rho} ) - (\delta_\xi
  \leftrightarrow \delta).
\end{equation}

\subsection{Reduction to the metric formulation}
\label{sec:reduction}

We now compare the expression \eqref{Final expression Pal} with the
standard results obtained in the metric formulation, where absence of
torsion and metricity are assumed. For this purpose, let us go
on-shell for the auxiliary fields ${\Gamma^\rho}_{\mu \nu}$ appearing
in the co-dimension 2 form \eqref{Final expression Pal}. One directly
gets
\begin{equation}
\begin{split}
  \kappa^{-1} k^{[\mu\nu]}_\xi=\sqrt{|g|} \Big[ &\xi^\sigma
  C^{\mu\nu}_{\;\;\;\sigma}-\xi^\mu
  C_{\;\;\sigma}^{\sigma\;\;\;\nu}+\half\xi^\mu
  C^{\nu\sigma}_{\;\;\;\sigma}-\half \xi^\nu D_\sigma
  h^{\mu\sigma}+\frac{1}{4}\xi^\nu D^\mu h\\ &- h^{\nu\sigma}
  D_\sigma \xi^\mu +\half h
  D^\nu\xi^\mu\Big]-(\mu\leftrightarrow\nu).
\end{split}
\end{equation}
When taking into account that $D_\mu$ is now the connection involving
the Christoffel symbols, so that
$C^\mu_{\;\;\tau\sigma}=\half(D_\tau h^\mu_\sigma + D_\sigma
h^\mu_\tau -D^\mu h_{\tau\sigma})$, we obtain
\begin{equation}
\kappa^{-1} k^{[\mu\nu]}_\xi=\sqrt{|g|}\left[\xi_\sigma D^\nu
  h^{\mu\sigma}+\xi^\nu D^\mu h -\xi^\nu D_\sigma
  h^{\mu\sigma}- h^{\nu\sigma} D_\sigma \xi^\mu + \half h
  D^\nu\xi^\mu\right]-(\mu\leftrightarrow\nu) . 
\end{equation}
Assuming that $\xi^\mu$ is a Killing vector, namely
$D_\sigma \xi^\mu + D^\mu \xi_\sigma = 0$, this expression reproduces
exactly the co-dimension 2 form obtained in metric formalism
\cite{Iyer:1994ys,Anderson:1996sc,Barnich:2001jy}. Therefore, we see
that the co-dimension $2$ forms expressions are equivalent in Palatini
and metric formalisms for exact reducibility parameters. However, this
result does not hold when $\xi^\mu$ is not a Killing vector. In
particular, the expressions may not match when using asymptotic
Killing vectors, as previously pointed out in the the Cartan
formulation of general relativity
\cite{Barnich:2016rwk,Oliveri:2019gvm}.

\section*{Acknowledgements}
\label{sec:acknowledgements}

\addcontentsline{toc}{section}{Acknowledgments}

This work is supported by the F.R.S.-FNRS Belgium through
conventions FRFC PDR T.1025.14 and IISN 4.4503.15. The work
of P.~Mao is supported in part by the National Natural Science Foundation
of China under Grant Nos. 11905156 and 11935009. The work of
R.~Ruzziconi is supported by a FRIA fellowship. 

\addcontentsline{toc}{section}{References}


\providecommand{\href}[2]{#2}\begingroup\raggedright\endgroup

\end{document}